\documentclass[twoside,11pt]{article}
\usepackage{mlhc}

\usepackage{wrapfig}
\usepackage{lipsum}
\usepackage[ruled]{algorithm2e}
\usepackage{algpseudocode}
\usepackage{bbold}
\usepackage{amsmath}
\usepackage[font=small,skip=2pt]{caption}

\newcommand{\pkg}[1]{{\fontseries{b}\selectfont #1}} 
\DeclareMathOperator*{\argmin}{arg\min}% Definitions of handy macros can go here

\ShortHeadings{Prediction Hierarchical Clustering}{Lorenzi, Brown, Sun, and Heller}
\begin{document}

\title{Predictive Hierarchical Clustering: \\ Learning clusters of CPT codes for improving surgical outcomes}

\author{\name Elizabeth C. Lorenzi\thanks{Authors contributed equally.}
 \email elizabeth.lorenzi@duke.edu \\
       \name Stephanie L. Brown$^*$ \email stephanielisa.brown@duke.edu \\
       \addr Department of Statistical Sciences\\
       Duke University, Durham, NC
       \AND
       \name Zhifei Sun \email zhifei.sun@duke.edu\\
       \addr Department of Surgery\\
       Duke University, Durham, NC
       \AND 
              \name Katherine Heller \email katherine.heller@duke.edu \\
       \addr Department of Statistical Sciences\\
       Duke University, Durham, NC
       }
\maketitle

\begin{abstract}
We develop a novel algorithm, Predictive Hierarchical Clustering (PHC), for agglomerative hierarchical clustering of current procedural terminology (CPT) codes.  Our predictive hierarchical clustering aims to cluster subgroups, not individual observations, found within our data, such that the clusters discovered result in optimal performance of a classification model. Therefore, merges are chosen based on a Bayesian hypothesis test, which chooses pairings of the subgroups that result in the best model fit, as measured by held out predictive likelihoods. We place a Dirichlet prior on the probability of merging clusters, allowing us to adjust the size and sparsity of clusters. The motivation is to predict patient-specific surgical outcomes using data from ACS NSQIP (American College of Surgeon's National Surgical Quality Improvement Program). An important predictor of surgical outcomes is the actual surgical procedure performed as described by a CPT code. We use PHC to cluster CPT codes, represented as subgroups, together in a way that enables us to better predict patient-specific outcomes compared to currently used clusters based on clinical judgment.
\end{abstract}
\section{Introduction}

With the widespread adoption of electronic health records (EHR) and the strong effort to digitize health care, there is a tremendous opportunity to learn from past experiences with patients to better adjust care for the current patient. Our goal is to use a surgical complications database (American College of Surgeon's National Surgical Quality Improvement Program) to train a model to predict post-operative surgical complications. Surgical complications, such as pneumonia, renal failure, and infection, are associated with decreased quality of life, inferior survival, and significant costs. Small improvements in prediction are meaningful in this application; major complications add substantial costs to the healthcare system, with an estimated \$11,500 of increased cost per event \citep{dimick2004hospital}.  To help align the interests of patients, providers, and payers, it is vitally important that health systems minimize the occurrence of any complication.

A major characteristic in predicting complications is which surgical procedure will be performed on the patient. The database contains 3,132 unique surgical codes, known as current procedural terminology, or CPT codes, along with 317 other covariates including patient specific information such as age, gender, lab values, and prior medical history. Clinical guidance suggests that there are underlying groups of surgeries which have different relationships between these predictors and the response of interest. Initial modeling via penalized logistic regression uses clinically formed groupings of CPT codes, which groups classes of similar surgeries together (e.g. general, cardiovascular, etc.). In this initial modeling, we fit separate models per grouping of CPT codes, resulting in different coefficients for each grouping. These CPT categories reduce the CPT codes to 16 clusters of procedures, however, these groupings are based on insurance and clinical information, and lack statistical learning behind the groups. In this paper, we develop a hierarchical clustering algorithm of surgical procedures (via CPT codes), with the goal of finding clusters that optimize the performance of a sparse logistic regression model for predicting surgical complications. 

Clustering is used to find underlying patterns and group structure by partitioning data into groups based on similarity between observations. Hierarchical clustering is one of the most frequently used unsupervised learning techniques, where the algorithm creates a binary tree (dendrogram) based on similarity between the data points \citep{duda1973pattern}. The result is a hierarchically formed structure which provides multiple clustering solutions, where often the hierarchy agrees with the intuitive organization of real-world data.

Many other clustering approaches exist that are in the form of hierarchically formed clusters. For example, Bayesian Hierarchical Clustering (BHC), is a probabilistic approach to hierarchical clustering that decides merges based on statistical hypothesis testing which compares the probability of the data belonging in a single cluster versus being separated based on marginal likelihoods \citep{heller2005bayesian}. Our approach adapts the ideas of this paper, where instead of comparing marginal likelihoods of covariates, we compare across predictive held out likelihoods based on learned penalized logistic regressions. Unlike BHC, we do not implement a fully Bayesian model. For computational purposes, we fit a lasso logistic regression using the R package, which greatly improves the speed of our algorithm compared to a Bayesian implementation of lasso with a logistic or probit link. 

We aim to learn clusters not based on similarity of data points, as learned through a marginal likelihood of a probabilistic model over data, but instead by which clusters, when merged, most improve the accuracy of prediction. This is done by comparing the probability that all data in a potential merge were generated from the same conditional distribution, $p(\mathbf{y}|\mathbf{X},\theta)$, to whether they were derived from separate models. In addition, we aim to cluster subgroups within our data; we approach this data challenge by initializing each cohort of patients with the same CPT code as a single cluster in our data. We learn a regression model for each potential merge on a training set using two-thirds of the data, then test each model on the held out third for evaluation in the testing set using predictive likelihoods under the learned regression model. The result is a dendrogram that hierarchically relates CPT codes based on which cohorts of patients improve the prediction of the overall model.  

%When implementing this algorithm we first split the data into a training set and a held-out validation set.  Then, using only that training set, at each iteration of the algorithm we learn a regression model on two-thirds or the training data and evaluate on the remaining one-third to determine each merge. 

\section{Algorithm}
\subsection{Overview}
Predictive hierarchical clustering (PHC) is similar to the traditional agglomerative clustering algorithm with its one-pass bottom-up approach that iteratively merges pairs of clusters. However, our clustering challenge involves additional data components, such as a response variable, \textrm{\textbf{y}}, a set of predictors, \textrm{\textbf{X}}, and nested subgroups within the data, \textrm{\textbf{z}}.  

% \begin{figure}[h!]
% \centering
% \includegraphics[width=.5\linewidth]{bhc_pic.png}
% \caption{Left: Portion of a tree where $\mathcal{T}_i$ and $\mathcal{T}_j$ are merged into $\mathcal{T}_k$, and the associated data sets $D_i$ and $\mathcal{D}_j$ are merged into $\mathcal{D}_k$. Right: Tree with 4 data points. The clusterings (1 2 3)(4) and (1 2)(3)(4) are tree-consistent partitions of this data. The clustering (1)(2 3)(4) is not a tree consistent partition.}
% \label{bhc}
% \end{figure}

Let $\mathcal{D} = \{(\mathbf{x}_1,{y}_1)..., (\mathbf{x}_n,{y}_n)\}$ denote the data set, and $\mathcal{D}_i \in \mathcal{D}$ the set of data points at the leaves of the subtree, $\mathcal{T}_i$. The data, $\mathcal{D}$, contains $G$ subgroups, representing the $G$ unique CPT codes. Each row of our data,$(\mathbf{x}_i,{y}_i)$, has a group assignment, $z_i = g$, where $g \in \{1,...,G\}$. The algorithm is initialized with $G$ trees, each containing all data for a single CPT procedural code, $\mathcal{D}_g = \{(\mathbf{X}_i,\mathbf{y}_i): \forall$ $i$ s.t. $z_i = g\}$. At each iteration, the algorithm considers merging all pairs of existing trees. If $\mathcal{T}_i$ and $\mathcal{T}_j$ are merged into the new tree $\mathcal{T}_k$, then the associated set of data for $\mathcal{T}_k$ is now $\mathcal{D}_k = \mathcal{D}_i \cup \mathcal{D}_j$. We restrict our search of merges to only partitions in the data consistent with the subtrees, $\mathcal{T}_i$ and $\mathcal{T}_j$. This reduces the number of comparisons at each potential merge, allowing for a more computationally efficient algorithm.

Two hypotheses are considered for each potential merge. The first hypothesis, denoted by $\mathcal{H}_1^k$, states that the data considered, $\mathcal{D}_k$, are generated from the same model, $p(\mathbf{y}_k|\mathbf{X}_k, \theta_k)$, with unknown parameter, $\theta_k$. We learn the model $p(\mathbf{y}_k|\mathbf{X}_k, \theta_k)$ using a penalized logistic regression over a training set of the data, $\mathcal{D}_{k,\textrm{train}}$. To evaluate the probability of the data under this hypothesis, $p(\mathcal{D}_k|\mathcal{H}_1^k)$, we compute a predictive likelihood on a held out set of data.
\begin{equation}
    p(\mathcal{D}_k|\mathcal{H}_1^k) = p(\mathbf{y}_{k,\textrm{test}}| g(\mathbf\mathbf{X}_{k,\textrm{test}}, \hat{\theta}_k))
    \label{hyp1}
\end{equation}
 
The predictive likelihood is solved by first estimating the predicted probability of an outcome under the learned logistic model, 
\begin{equation}
    \hat{\mathbf{p}} = g(\mathbf\mathbf{X}_{k,\textrm{test}}, \hat{\theta}_k) =  p(\mathbf{y}_{k,\textrm{test}}=1|\mathbf{X}_{k,\textrm{test}}, \hat{\theta}_k) = \frac{1}{1+\exp{\{-\mathbf{X}_{k,\textrm{test}} \hat{\theta}_k\}}}
\end{equation}
where $\hat{\theta}_k$ are the learned coefficients from the model. We then evaluate the likelihood of our data under those predicted probabilities, $p(\mathbf{y}_{k,\textrm{test}}|\hat{\mathbf{p}}) = \prod_{i\in k}^{n_k} \text{Bernoulli}(\mathbf{y}_{k,\textrm{test}}, \mathbf{\hat{p}})$. Though our application works with a binary response, this is easily adapted for continuous response data using a normal likelihood and solving first for the predicted mean instead of $\hat{\mathbf{p}}$.

The second hypothesis, denoted by $\mathcal{H}_2^k$, states that the data considered, $\mathcal{D}_k = \{\mathcal{D}_i \cup \mathcal{D}_j\}$, are generated from separate models with different parameters $\{\theta_i,\theta_j\}$.  When evaluating the second hypothesis, we are always considering clusterings that partition the data in a manner consistent with the subtrees, $\mathcal{T}_i$ and $\mathcal{T}_j$. This allows us to avoid summing over the exponential number of possible ways of dividing the data, $\mathcal{D}_k$, into two clusters, and allows us to utilize recursion in evaluating the probability of the two data sets, $\mathcal{D}_i$, $\mathcal{D}_j$, under each previously merged tree,  $p(\mathcal{D}_i|\mathcal{T}_i)$ and $p(\mathcal{D}_j|\mathcal{T}_j)$ (defined in Equation (\ref{pd_t})). The probability of the data under the second hypothesis is $p(\mathcal{D}_k|\mathcal{H}_2^k) = p(\mathcal{D}_i|\mathcal{T}_i)p(\mathcal{D}_j|\mathcal{T}_j)$. To evaluate the probabiilty of the data under each tree, we combine the probability of the data under both hypotheses and weight by a prior probability that all points belong to one cluster, $\pi_k = p(\mathcal{H}_1^k)$. 

\begin{equation}
p(\mathcal{D}_k|\mathcal{T}_k) = \pi_k p(\mathcal{D}_k|\mathcal{H}_1^k) + (1-\pi_k) p(\mathcal{D}_i|\mathcal{T}_i)p(\mathcal{D}_j|\mathcal{T}_j)
    \label{pd_t}
\end{equation}

For each possible merge we first compute these hypotheses, then choose our next merge based on which trees, $\{\mathcal{T}_i,\mathcal{T}_j\}$ when merged into a single tree $\mathcal{T}_k$ result in the largest improvement in $p(\mathbf{y}_k|g(\mathbf{X}_k, \hat{\theta_k}))$ compared to $p(\mathbf{y}_i|g(\mathbf{X}_i, \hat{\theta}_i))p(\mathbf{y}_j|g(\mathbf{X}_j, \hat{\theta}_j))$. We evaluate this improvement using Bayes rule, where we compute $r_k = p(\mathcal{H}_1^k|\mathcal{D}_k)$, shown in Equation (\ref{compare_hypothesis}).
\begin{equation}
r_k = \frac{\pi_k p(\mathcal{D}_k|\mathcal{H}_1^k) }{\pi_k p(\mathcal{D}_k|\mathcal{H}_1^k) +(1-\pi_k)p(\mathcal{D}_k|\mathcal{H}_2^k)}
\label{compare_hypothesis}
\end{equation}

The two hypothesis effectively test that the association of the predictors with a surgical outcome variable, $\mathbf{y}$, is the same for considered data, $\mathcal{D}_k = \{\mathcal{D}_i \cup \mathcal{D}_j\}$, versus allowing the association to be specific to each tree under consideration. The formal algorithm is presented in Algorithm \ref{algo1}.

\begin{algorithm}
\small
\caption{Predictive Hierarchical Clustering}
    \KwData{Response: $\mathbf{y} \in \{y_1,...,y_n\}$ \newline Predictors: $\mathbf{X} \in \{\mathbf{x}_1,...,\mathbf{x}_n\}$ 
    \newline Group Membership: $\mathbf{z} = \{z_1,...,z_n \}$ where $z_i \in \{1,...,G\}$} 
    \SetKw{KwInit}{Initialize:} 
    \KwInit{Number of clusters $c = G$
    \newline
    \hspace*{1.7cm}$\mathcal{D}_g = \{(\textbf{x}_i,y_i ): z_i = g\}$ \newline
    \hspace*{1.7cm}Fit model $\mathcal{D}_{g,\textrm{train}}$ $\forall$ $g$ \newline 
    \hspace*{1.7cm}Save $p(\mathcal{D}_{g,\textrm{test}}|\mathcal{H}_1^g)$ $\forall$ $g$ \newline}
    \While{$c>1$}{
    Find the pair $\mathcal{D}_i$ and  $\mathcal{D}_j$ with the highest probability of the merged hypothesis: \newline
    \[r_k = \frac{\pi_k p(\mathcal{D}_k|\mathcal{H}_1^k) }{\pi_k p(\mathcal{D}_k|\mathcal{H}_1^k) +(1-\pi_k)p(\mathcal{D}_k|\mathcal{H}_2^k)}\]
    Using the following steps:
    \begin{itemize}
    \item $\forall \enspace  \mathcal{D}_k = \mathcal{D}_i \cup \mathcal{D}_j$ s.t. $i \neq j$, fit $p(\mathbf{y}_{k, \textrm{train}}|\mathbf{X}_{k,\textrm{train}}, \theta_k)$ to learn $\hat{\theta}_k$
    \item Predict on test set of data using learned coefficients: $\hat{\mathbf{p}} = p(\mathbf{y}_{k,\textrm{test}}=1|\mathbf{X}_{k,\textrm{test}}, \hat{\theta}_k)$
    \item Calculate likelihood of test data under predicted probability, $p(\mathbf{y}_{k,\textrm{test}}|\hat{\mathbf{p}})$, to learn $p(\mathcal{D}_k|\mathcal{H}_1^k)$
    \item Use previously stored $r_k$ to evaluate $p(\mathcal{D}_k|\mathcal{H}_2^k)$
    \item Solve for $r_k$
    \end{itemize}
    Merge $\mathcal{D}_k \leftarrow \mathcal{D}_i \cup \mathcal{D}_j$ \\
    Delete $\mathcal{D}_i$ and $\mathcal{D}_j$ \\
    c= c-1
    }
\label{algo1}
\end{algorithm}

\subsection{Implementation}
%this sounds a little awkward as a transition but i think how we describe the models is useful for clarity
The model is initialized by running a regression for each subset of data, $\mathcal{D}_g$. We then consider all possible merges, for example merging trees $\mathcal{T}_s$ and $\mathcal{T}_t$ into $\mathcal{T}_u$, by running a model on the subset of data where $z_i = s$ combined with the data where $z_i = t$. The regression model for this application is the lasso logistic regression using all covariates as main effects with no interactions, penalizing the complexity of the models with the lasso penalty. We choose the lasso penalization for its ability to handle strongly correlated covariates and to perform variable selection \citep{tibshirani1996regression}. This method was preferred over an assortment of commonly used binary classifiers based on its predictive performance and computational efficiency. The logistic regression equation with the incorporated penalty is shown in (\ref{1}). Any type of classification or regression model is usable in this algorithm, making the framework flexible for many different data problems.

% \begin{equation}
% \textrm{logit}(p(\mathbf{y}_i=1|\mathbf{X}_i, \theta_u)) = \mathbf{X}_i\theta_{u} + \epsilon_i, \forall i: z_i = s \lor t, \quad s \cup t = u
% \label{mergemod1}
% \end{equation}

\begin{equation}
L(\mathbf{\beta}) = \argmin_{\mathbf{\beta}} \sum_{i=1}^n y_i \mathbf{x}_i\beta -\log(1+\exp{(\mathbf{x}_i \mathbf{\beta})}) + \lambda_1\|\mathbf{\beta}\|_1
\label{1}
\end{equation}

Because the algorithm requires the fitting of multiple regressions at each iteration of the tree, our algorithm is constructed in a way that allows parallelization. The complexity of computing the tree is $O(n^2)$, however, as each hypothesis comparison is independent of all others we can compute the merge metric for multiple considered merges simultaneously.  At each iteration of the algorithm, we push the possible models to learn to separate cores of the machine to reduce the computational time by a factor of the number of available cores. 

Once the hierarchical tree is learned, we cut the tree to retrieve cluster solutions. The merges in the tree are made at the chosen $r_k$ for that iteration of the algorithm, resulting in the merge being plotted in the dendrogram at a height equal to $r_k$. When the weighted probability of the merged hypothesis, $r_k$ $>0.5$, the merges are justified. Therefore, the tree should be cut at points where $r_k < 0.5$.

  \begin{wrapfigure}{r}{0.4\textwidth}
    \centering
    \vspace{-18pt}
    \hspace{5pt}
      \begin{algorithm}[H]
\small
\caption{Updating $\pi_k$}
    \SetKw{KwInit}{Initialize:} 
    \KwInit{Each node of tree $i$ to have $d_i=\alpha$, $\pi_i=1$ \newline}
    \For{each internal node $k$}{
        \[d_k = \alpha \Gamma(n_k) + d_{\text{left}_k}d_{\text{right}_k}\]
        \[\pi_k = \frac{\alpha \Gamma(n_k)}{d_k} \]
    }
\label{algo2}
\end{algorithm}
  \end{wrapfigure}
  
\subsection{Learning $\pi_k$}
  
To learn an appropriate $\pi_k$, or the prior probability of the merged hypothesis, a Dirichlet prior is placed on this parameter and updated at each step of our algorithm. The approach is similar to Dirichlet Process mixture models (DPM), where the probability of a new data point belonging to a cluster is proportional to the number of points already in that cluster \citep{blackwell1973ferguson}. This is commonly modeled with a Dirichlet prior (for finite models) or a Dirichlet Process prior (for infinite models) over the sampling proportions of each cluster with a concentration parameter, $\alpha$, that controls the expected number of clusters. Although we are not approximating a Dirichlet process mixture model, the prior over $\pi_k$ represents the probability of merging over all the possible partitions in the tree. The resulting prior update on the merged hypothesis calculates the relative mass of all subgroups, $z_k$, belonging to one cluster versus the subgroups, $z_k$, remaining in their current partition as defined by the tree structure. This process is clearly defined in Algorithm \ref{algo2}, where we denote the right subtree of $\mathcal{T}_k$ as $\text{right}_k$ the left subtree as $\text{left}_k$, and the number of subgroups within the tree as $n_k$.

\section{Results}

\subsection{Simulated Results}
To validate the PHC algorithm, we simulate data which inherently is structured with subgroups and verify that the subgroups of data are merged appropriately. We first simulate four sets of coefficients, $\{\theta_1, \theta_2, \theta_3, \theta_4\}$ drawn from a standard normal, and next create $\mathbf{X}$ from standard normal draws. We convert some of the variables in $\mathbf{X}$ to binary variables (using a normal CDF) to be as representative of our data as possible. Next, we compute the associated $\mathbf{y}$ with twenty subgroups, drawn from a Bernoulli distribution with the probability of the outcome being the inverse logit of $X\theta$, where the first set of five subgroups use $\theta_1$, the second five subgroups use $\theta_2$, etc.; therefore giving the inherent clustering of the twenty subgroups into four larger clusters.  For simplicity, subgroups 1-5 are generated from one distribution of coefficients, 6-10 another, 11-15 a third, and 16-20 the fourth.
\begin{figure}[h!!]
\centering
\includegraphics[width=.8\linewidth, height=2.2in]{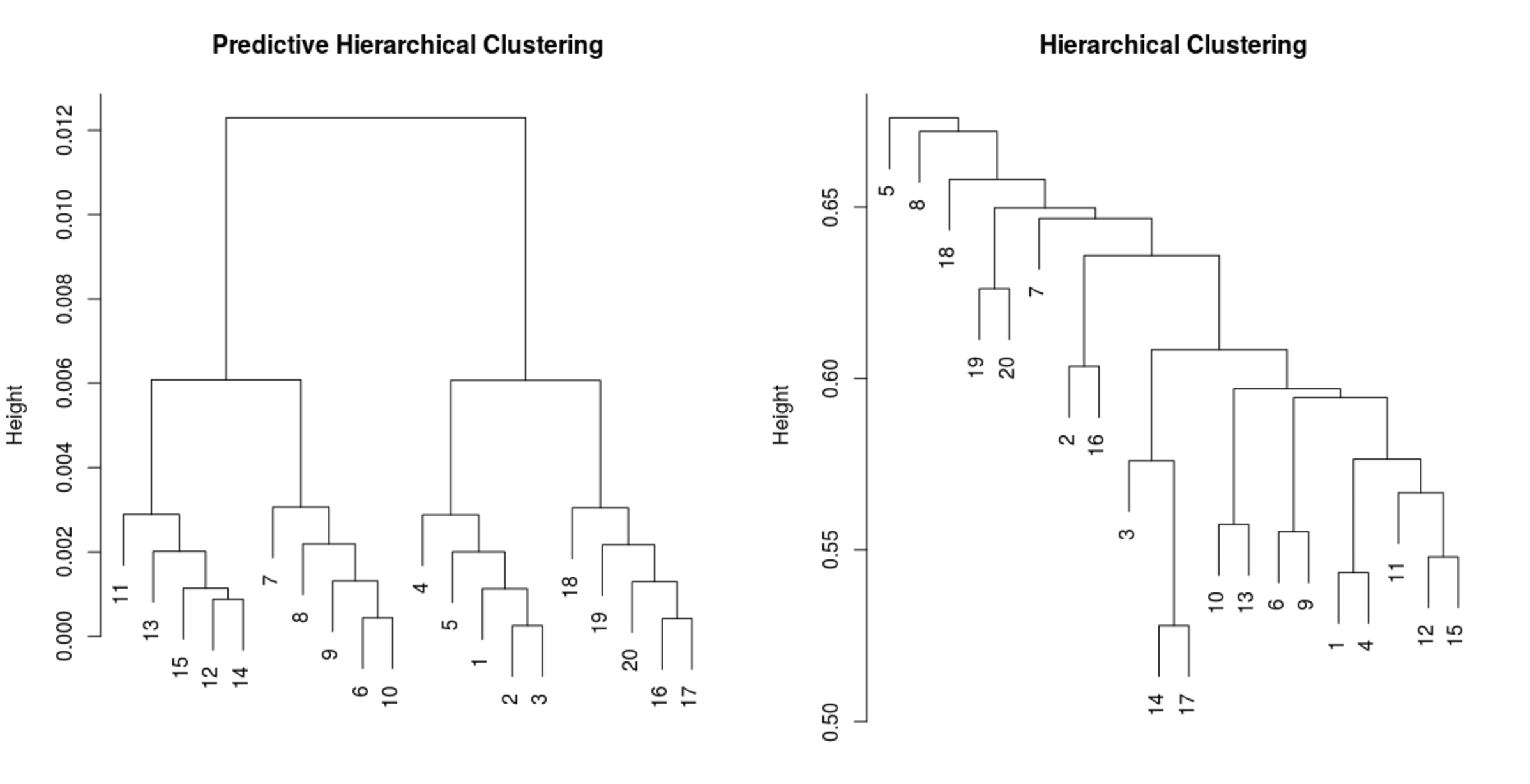}
\caption{ Clusters discovered by PHC (left)  and clusters discovered by traditional hierarchical clustering with complete linkage (right). Evaluated on simulated data with twenty subgroups and four underlying groups.}
\label{int_dend}
\end{figure}
The dendrogram in Figure \ref{int_dend} left confirms that we are able to find the underlying structure in the data. As a comparison to other methods, Figure \ref{int_dend} right displays the results from running traditional hierarchical clustering with complete linkage, using the R package, \pkg{hclust}. Performing traditional hierarchical clustering on this simulated data with $n$ observations falling into $G$ subgroups would result in a dendrogram with $n$ nodes instead of $G$. Therefore, we average over each predictors for each subgroup, and then calculate Euclidean distance between these points. The resulting \pkg{hclust} dendrogram is unable to separate the data into the four true subgroups when clustering based solely on similarity.

\begin{figure}[h!!]
\centering
\includegraphics[width=.8\linewidth, height=2.3in]{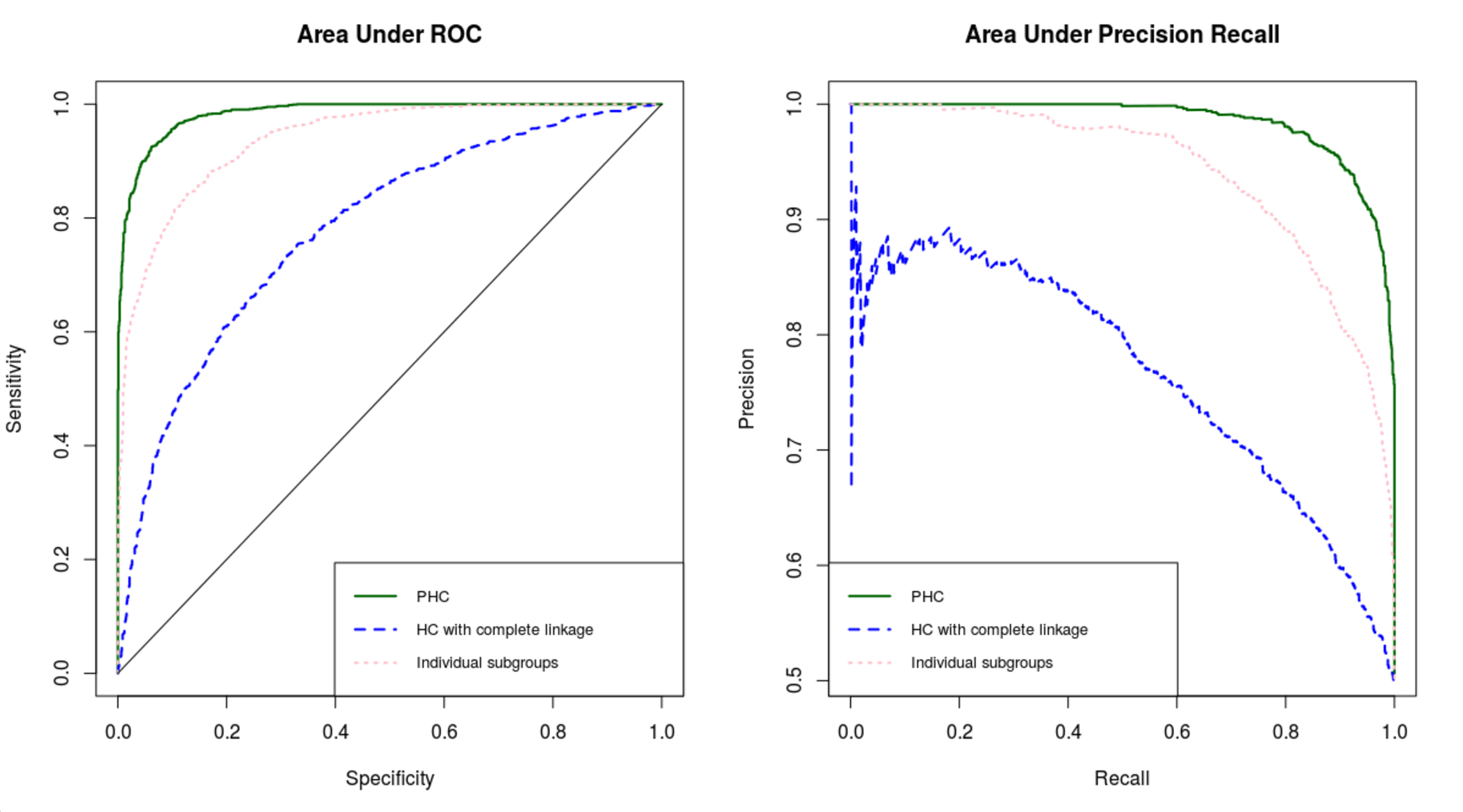}
\caption{ROC curve (left) and precision recall curve (right) from simulated data showing clustering solutions for PHC (dark green) compared to hierarchical clustering with average linkage (blue) and treating each subgroup as an individual clulster (pink). Data corresponds to that used in Figure \ref{int_dend}.}
\label{roc_slope}
\end{figure}

We additionally are interested in whether these new clusters improve the overall prediction of the penalized logistic regression. We test the clusters by fitting models for each of the four groups found by PHC and compare these to cutting the \pkg{hclust} dendrogram at four clusters then fitting models. As a further comparison, we fit models for each of the 20 subgroups individually. The data is split into a training set to fit the models and a testing set to generate predictions.  We plot the ROC curves and compare the results. As can be seen in Figure \ref{roc_slope}, the four clusters found by PHC substantially outperform the traditional hierarchical clustering approach and the twenty subgroups modeled individually. 

\subsection{Experimental Results}
The development of PHC is motivated by our goal of accurately predicting the risk of adverse outcomes for individual patients while incorporating procedural information in an informed way.  We work with data from American College of Surgeons' NSQIP, a national surgery program collected from participating hospitals across the country. The data contain information about surgery patients' demographic, medical history, lab results, procedural information via CPT codes, diagnosis information via Clinical Classifications Software (CCS) categories, and postoperative outcomes.

Specifically, we are working with 317 predictor variables not including the CPT information.  These predictors include continuous variables, such as lab values, as well as binary indicators of patient history, such as whether a patient has diabetes. Many of the indicator variables are sparse, with the majority of indicators with means less than $5\%$. Some of the continuous variables are highly skewed, so a normal scores transform is applied to handle the non-normality \citep{bogner2012technical}.  In addition, we have diagnosis information grouped through CCS categories, providing context to which ailment the patients have. The six post-operative outcomes are any morbidity (any type of complication), surgical site infection (SSI), respiratory complication, septic, 30-day mortality, and renal failure. We use a subset from NSQIP with data spanning from 2005-2014. This subset of the data is a sample of 3,723,252 patients, with 3,132 unique CPT codes. Each patient has one unique surgical CPT code describing the main surgical procedure performed.
 
%Previous clusterings of CPT codes are provided by Clinical Classifications Software (CCS).  This system was developed at the Agency for Healthcare Research and Quality (AHRQ), and is a tool for clustering patient diagnoses and procedures into a manageable number of clinically meaningful categories \citep{healthcare2011clinical}.  Although more manageable than 2,790 unique codes, there are still 138 of these groupings in our data and additionally they were determined based on clinical relevance as opposed to statistical similarity between procedures.  We use these categories as a baseline to compare model performance after PHC clusters are determined.

CPT codes naturally fall into a hierarchical system where there exist coarse groupings of CPT codes based on the type of surgery performed.  For example, CPT codes in the range of 30000:32999 are all surgeries performed on respiratory system, codes in the range of 40490:49999 are digestive system surgeries, and others fall into categories for cardiovascular system, urinary system, and nervous system amongst others \citep{american2017current}.  There are a total of 16 categories a code may fall in.  We use these categories as a baseline to compare to our hierarchically learned groupings. In addition, we compare to using all CPT codes (without being clustered) to assure that the learned clusterings do not lose information that could be gained from using the data directly.

%As these predetermined categories were not chosen with the objective of improving prediction of surgical outcomes we suspect that our PHC learned groupings give improvements on prediction.

The number of patients with the same CPT code varies greatly. Some procedures only occur once or twice in our data while other more common procedures occur tens of thousands of times.  In addition, the proportion of adverse outcomes varies between CPT codes and is often less than 2\%;  in Table \ref{table}, we show the prevalence of each outcome across all CPT codes. As there is little information for a model to learn on small groups with few realizations of adverse outcomes, we restrict the number of observations per CPT code to be included in our model.  For model stability we choose groups with at least 500 patients. This allows 646 of the 3132 possible CPT codes to be considered, but the excluded CPT codes are so small that the ones included encompass 94\% of the data available. Large subgroups are needed due to the sparsity of outcomes in combination with the need to split the data into train (learn each model), test (calculate predictive likelihood), and validation sets (evaluate performance of clusters outside of algorithm), and the need for the training set to be split further for cross-validation of the lasso shrinkage parameter.

After running PHC independently for each outcome, we observe the resulting dendrogram and decide where to cut the tree as guided by the predictive likelihoods from the output of the tree, as discussed in the previous section. To verify that the PHC clusters improve the model over the baselines, we compute the area under the ROC curve and precision recall using held-out validation data.  We use baselines of Lasso models fit using the same structure but with different groupings of CPT codes; unique random slopes for each PHC cluster, clinical cluster, or CPT code.  The results are shown in Table \ref{table}, where we see improvement for the majority of the outcomes using PHC. Specifically, our learned clusters improve the AUC for the outcome, overall morbidity, by 3.8\% compared to the baselines and more than 1\% for surgical site infection (SSI) and respiratory outcomes. The margin of improvement from PHC over the baselines decreases for the sparser outcomes. Because our algorithm fits separate models for each CPT code initially, if that CPT code has fewer patients and fewer instances of that outcome, then the model fit may suffer and therefore the algorithm has trouble merging these groups. In final testing, once clusters are learned, it is common for the clusters encompassing the most patients to perform best. Additionally, the models trained on the clinical groups often perform better than the clusters per CPT models.  There is a clear indication that larger training samples in each cluster improves model fit.

\begin{table*}[t]
\small
\centering
\begin{tabular}{|c|c|c|c|}
\hline
 \textbf{Outcome} & \textbf{Prevalence}& \textbf{AUROC}  & \textbf{AUPRC }\\ 
  & & (PHC/Clinic/All) & (PHC/Clinic/All)\\ \hline
   Morbidity & 14.51\% & \textbf{0.839}/0.801/0.798 &  \textbf{0.685}/0.600/0.605\\ 
   SSI & 4.13\% & \textbf{ 0.671}/0.655/0.651 &  \textbf{0.209}/0.191/0.184\\ 
   Respiratory & 2.77\% & \textbf{ 0.845}/0.833/0.810 &  \textbf{0.410}/0.407/0.381\\ 
      Septic & 2.57\% & \textbf{0.740}/0.738/0.728 &  \textbf{0.292}/0.274/0.289\\ 

   %UTI & 1.53\% & \textbf{ 0.647}/0.646/0.605 &   0.068/0.068/0.060\\ 
   Mortality & 1.27\% & \textbf{ 0.869}/0.861/0.835 &  \textbf{0.381}/0.356/0.293\\ 
   Renal& 0.65\% & \textbf{ 0.808}/0.801/0.804&   \textbf{0.157}/0.137/0.149\\ 
 \hline
\end{tabular}
\caption{Prevalence of each outcome, along with area under receiver operator curve (AUROC) and area under precision recall curve (AUPRC) as found by regressions using the PHC clusters, predetermined course CPT groupings and a regression with each CPT in its own group. We denote the 16 clinically formed clusters of CPT codes as ``Clinic'' in the table.}
\label{table}
\end{table*}

Though the improvements for some outcomes are marginal, in an application of predicting surgical outcomes, where costs are high and lives are at stake, even small improvements can make a meaningful difference. Beyond the accuracy of prediction, the model parameters are also impactful for future clinical decision making. Learning clusters that are based on improved model fit, provides better estimates of coefficients that better guide clinicians to the patient's characteristics that are most attributing to their risk. 

In order to better assist doctors, we developed a software platform to display the predicted risk of complications and suggest interventions based on the predictors that are most influential in the inflated risk for a given patient. Our resulting clusters each have learned coefficients describing the relationship between the predictors and the outcome. Using the many-to-one mapping between the clusters and CPT codes, we can efficiently provide personalized results for the patient based on the surgical procedure the patient will undergo. For example, if the model learns that high levels of a certain lab value are associated with increased risk of renal failure after a specific procedure and that patient shows an abnormally high value of that lab, clinicians can be alerted and intervene with appropriate treatment to help mitigate the risk of that outcome.  As the influence of each predictor varies across clusters it is essential that we correctly determine clusters for each procedure. The interpretability of the model provides tools for clinical interventions to prevent complications.

We aim to test feasibility and measure the benefit of using this software in actual practice on a surgical ward. Validation of the software will support its adoption in daily medical practice, both at our institution and elsewhere. The interface will eventually be linked to the health system's EHR, making it an easy transition into the daily work flow of health care providers. This is in contrast to the ACS risk calculator that requires manual data input prior to any risk calculations being obtained \citep{bilimoria2013}.  

\section{Related Work}
Prediction and clustering are two very commonly used data mining techniques. In \citet{blockeel2000top}, they present a method that adapts decision trees to the task of clustering for prediction by employing instance-based learning. Similarly, in \citet{vzenko2005learning}, they introduce predictive clustering rules, where each clustering can be considered as a ``rule" that defines that cluster. Another realm of clustering is "clusterwise linear regression", first introduced by \citet{spath1979algorithm}, who proposed an algorithm that forms K partitions with corresponding sets of parameters that minimize the sum of errors across each cluster. In \citet{desarbo1988maximum}, they present a conditional mixture, maximum likelihood approach to learn clusterwise linear regression. The downside to these clusterwise linear regression approaches include requiring the number of clusters and initial partitions to be pre-specified, and the proclivity of getting stuck in local modes. 

Many other clustering approaches exist that are in the form of hierarchically learned clusters. For example, we have discussed how PHC relates to Bayesian Hierarchical Clustering (BHC),  a probabilistic approach based on marginal likelihood calculations \citep{heller2005bayesian}. Similarly, \citet{neal2001defining} uses Dirichlet diffusion trees to provide a probabilistic hierarchical clustering approach. The work of \citet{teh2009bayesian}, Bayesian Agglomerative Clustering with Coalescents, is a fully Bayesian approach for hierarchical clustering based on a prior over trees called Kingman's coalescent. Other similar approaches can be seen in \citep{williams1999mcmc}, \citep{kemp2004semi},\citep{roy2007learning}.

The cited works have proposed predictive clustering techniques, however, none have learned clusters of nested subgroups in the data to improve overall prediction. The importance of clustering subgroups in our work is essential for future predictions. By learning a many-to-one mapping of CPT codes to clusters, we immediately can evaluate a current patient's risk without requiring updated clustering. The shared disadvantage of the mentioned probabilistic hierarchical clustering algorithms is their neglect of improving predictive performance of a model.

\section{Discussion}
We present a novel algorithm for predictive hierarchical clustering, where the end result are clusters of CPT codes which improve prediction of a regression model. Though our method is inspired by surgical complications data, PHC is very applicable in other applications. The algorithm has the advantage of transforming thousands of nested subgroups in data into larger clusters that better inform the models' predictive capabilities.  Clustering has proven useful in countless applications to group data by similarity, and the added power of grouping to improve prediction will be crucial in many applications. 

A current limitation of PHC is its inability to work with subgroups that contain only small amounts of sparse data.  With sparse outcomes and minimal signal we were forced to restrict the CPT groups used in the algorithm to those containing 500 or more observations. In addition, even with subgroups of that size, some CPT codes have more heterogeneous patient populations which results in models with little to no signal (in Lasso, this is seen through all variables being zeroed out). This is expected when considering this application. For example, a CPT code for gallbladder removal most likely encompasses a large patient population with very different characteristics making it difficult for a model to relate patient covariates to an outcome. Whereas, open heart surgery tends to be performed on a more specific type of patient, so the model can better learn the relationship between those patients and an outcome. A further limitation is the runtime, which is quadratic in the number of subgroups.  By eventually exploiting a randomized version of the algorithm, we can decrease the runtime to O(nlogn).

Our overarching goal of this work is to expand and test a Clinical Analytic \& Learning Platform in Surgical Outcomes (CALYPSO). Our objective is to develop an integrated computing platform for (1) data intake from the EHR, and (2) prediction of surgical risk using the described predictive models.  The user-interface incorporates existing data from the EHR and estimates an individual patient’s relative and absolute risk profile. We also display the most important and modifiable predictors via our user-interface based on the patient's profile and model coefficients. Each flagged predictor is then linked to a set of accepted best practice interventions that specifically targets the risk factor. Visualizing these individualized risks and interventions at the point-of-care allows clinicians to make data-driven decisions rapidly.

%APPENDICES are optional
%\balancecolumns
\bibliography{sample}

\end{document}